\documentclass[aps,prb,preprint,showpacs,showkeys,groupedaddress]{revtex4}
\usepackage{graphicx}
\usepackage{amssymb,amsmath}

\begin{document}

\title{Convergence method for calculating solutions to the 3D invariant embedding integro-differential equations describing electron transport processes.}

\author{C. Figueroa}
 \affiliation{Laboratorio de F\'{i}sica del S\'{o}lido,  Dpto. de
F\'{i}sica, Facultad de Ciencias Exactas y Tecnolog\'{i}a,
Universidad Nacional de Tucum\'{a}n, Argentina.}

\author{H. Brizuela}
 \affiliation{Laboratorio de F\'{i}sica del S\'{o}lido,  Dpto. de
F\'{i}sica, Facultad de Ciencias Exactas y Tecnolog\'{i}a,
Universidad Nacional de Tucum\'{a}n, Argentina.}

\author{S. P. Heluani}
\affiliation{Laboratorio de F\'{i}sica del S\'{o}lido,  Dpto. de
F\'{i}sica, Facultad de Ciencias Exactas y Tecnolog\'{i}a,
Universidad Nacional de Tucum\'{a}n, Argentina.}

\date{\today}

\begin{abstract}
The electron and photon transport processes in spectroscopy techniques described by the invariant embedding theory is here revisited. We report a convergence method to obtain closed analytical solutions to the 3D integro-differential equations.  This method was successfully used in calculating the dependence of the electron backscattered fraction on the atomic number and on the energy. Also the fraction of absorbed electron as a function of incident angles was calculated. Using a states ladder model for the electron energies, this method provides a tool for testing physical parameters involved in the transport theory, such as the elastic and inelastic cross sections.
The outstanding feature of the invariant  embedding differential equations of considering  observable quantities (such as the emergent flux of particles) as independent variables makes  them a suitable tool to describe experimental situations.

\end{abstract}

\pacs{68.49-h, 78.70.-g}

\keywords{invariant embedding, electron backscattering, electron spectroscopy}

\maketitle

\section{\label{sec:level1}Introduction}
Advances in characterization techniques applicable to the complex structures of materials such as nanorods, thin films, nanofiltration membranes etc., will advance the understanding of the relationship between structure and properties in the search for technological applications.  In the field of structural and chemical characterization methods, the electron and photon transport has several applications including Electron Probe Microanalysis (EPMA), X-ray Fluorescence Analysis, Electron-Beam-Induced-Current, Auger Electron Spectroscopy, Electron Energy Loss Spectroscopy, X-ray Photoelectron Spectroscopy (XPS), and Secondary Electron Emission. All these characterization techniques are based on the strong interaction of electrons and/or photons with matter and in the emissions or 'signals' that they produce. The different applications are defined according to which of these ´signals´ are detected and how they are analyzed. Particularly, in quantitative microanalysis, the energies and intensities of the characteristic x-ray lines from an irradiated sample are used to determine the elemental composition or dimension of small samples and also the films thicknesses. To quantify it the `signal' must be  corrected using some transport model and also some `correction model'. The loss of ionization due to backscattered electrons is one of the aspects that the correction models have to take into account and an example of the complexity of the problem. Therefore, most of the approaches which form the basis of the correction procedures commercially available in spectroscopy techniques, are still empirical or semi-empirical ones \cite{Heluani:XRS05}.

 Theoretical and experimental studies of the backscattered electrons are valuable in several techniques, such as reflection Electron Energy Loss Spectroscopy, Particle Induced X-ray Emission and Elastic Peak Electron Spectroscopy. Also their applications in Scanning Electron Microscopy (SEM) and in film thickness determinations are of great importance. In particular, SEM represents a high-performance method for investigating structures and devices in the nanometer scale\cite{Wang:AM03,Yao:05,Vita:RJP04}.

To deduce the presence of a particular chemical element, or to measure the thickness of a film or the size of a particle from spectroscopy measurements, it is necessary to describe the  emergent flux of particles (an observable quantity) in terms of the interactions produced by the   injected flux  during its trajectory inside a solid sample. Several  mathematical models were developed to describe the evolution of the electrons or photons in a solid medium. The application of the Boltzmann equation, or similar transport equations, presents serious difficulties  in solving the resulting integro-differential equations and little progress has been made in the solution of 2D and 3D transport problems. In addition more difficulties arise in the practical applications of theoretical results to different techniques \cite{Bernasconi:PRB88, Jablonski:JVST03, Werner:PRB05,   Werner:SIA05}. For example, Bernasconi  et al. \cite{Bernasconi:PRB88} consider the hot-electron transport in the conduction band of a thin insulating film in terms of the energy-dependent elastic and inelastic scattering rates. The authors obtain an exact formal solution for a system of integro-differential Boltzmann equations only for the zero-energy-loss transport problem. Werner \cite{Werner:PRB05} considered the backscattering of medium energy electrons from solid surfaces analyzing a linearized Boltzmann-like kinetic equation. Then, an approximate solution is implemented in a Monte Carlo scheme.

The Boltzmann transport equation and Monte Carlo calculations (which also make use of the Boltzmann transport equation to simulate the track of photons or electrons) are still  the most used methods for the interpretation of the signals in spectroscopic and characterization techniques. However, they are uneconomical methods, since much of the information obtained about the internal fluxes is quite useless to the experimenter. More rigorous treatments reflexing the complexities of the problems are the semiclassical treatments and Fresnel optics equations \cite{Dashen:PRB64, Vicanek:Ssci99, Lennard:NIM02} and quantum treatments \cite{Denton:PRA98}.

In recent articles the so-called Invariant embedding Approach to Microanalysis (IIAM) was developed  with the aim of obtaining analytical solutions to the transport problems associated to microanalysis techniques. The IIAM is free of the drawbacks of the empirical methods and the mathematical difficulties to solve the Boltzmann equations (see ref.\cite{Heluani:XRS05}). Using a simple 1D model, expressions for the detected X-rays characteristic intensities had shown a good performance in the interpretation of experimental data.\cite{Heluani:NIM02, Heluani:JAP06}.\\

The invariant embedding method is a mathematical technique based directly upon the physical processes. In this frame,  the traditional  linear functional equations, subject to boundary conditions, are transformed  into nonlinear functional equations subject only to initial conditions in space and time coordinates. Although analytical approaches to solve invariant embedding differential equations were reported in different physics contexts, the method has, as yet, had relatively little impact in spectroscopy techniques \cite{Ambarzumian:58, Chandrasekhar:Dov60, Bellman:JMP60, Wing:63,Kim:PRB98,Glazov:NIMB07,Kim:PRB08}.\\
  The purpose of this article is to report a convergence method for deriving an analytical solution to the 3D invariant embedding equations associated to a states ladder model for electrons transport processes within a solid medium.
 Considering the convergence of an approximate solution to the exact solution in a medium of infinite size, the distributions of the backscattered and the absorbed electron  fractions are obtained. In section II a brief description of the geometrical configuration and the parameters used in the theory is presented. Then, analytical expressions for the electron fluxes are obtained as functions of physical and experimental parameters such as the ratio between the elastic and inelastic cross  sections, the incident energy and the incident and emergent angles. Finally, the theoretical results are compared with experimental data.

  \section{Method of Convergence.}
 \subsection{Parameters of the theory.}

Considering a typical experimental design in EPMA, one of the  advantages  of the invariant embedding method is to take the sample thickness as the integration variable in the differential transport equations (instead of the coordinate measured from the sample surface, as  usual in the traditional Boltzmann equation treatment). Then, it is helpful to keep in mind the fact that the theoretical expressions for the different fluxes of particles (such as backscattered electrons) are written as a function of the coordinate  measured from the right end of the sample which is also the thickness of the sample, see figure 1b.

Let us  consider an usual experimental configuration in quantitative electron spectroscopy methods, as sketched in Figure 1-b, where a solid sample of a finite thickness $\tau$ is irradiated from the left by a beam of electrons with energy $E_{0}$ (this geometrical configuration could be applied to other similar techniques and the {\it electrons} could be {\it photons}). Some of the impinging electrons of the primary beam could be backscattered, some others could  become {\it deactivated} (i.e. they become unable to produce ionizations or characteristic x-rays inside the sample) and some others could escape from the other side (in the case of film samples or  light particles). In passing through the material the electrons undergo elastic and inelastic collisions. Both interactions are described by cross sections that rigorously are energy and angle dependent.
In this work we consider the usual axial symmetry and only azimuthal integrated cross sections is considered.

To clarify the notation, the diagrams in Figure 1a shows the angles $(\alpha_{j}$ and $\beta_{j})$, which are the  incident and emerging polar angles for a single interaction, and the index $j=0,1,..$ labels the distinct energy states values. The polar angles are measured with respect to an axis perpendicular to the sample surface.

       For isotropic elastic cross sections, the integral with respect to the azimuthal angle leads to,
    \begin{equation}
               d\sigma_{\beta_{j}}=2\pi \ k \ sin\beta_{j} d\beta_{j}. \nonumber
 \end{equation}
Where k is a constant. The integrated cross section is: $\sigma_{j}=\int d\sigma_{\beta_{j}}= 4 \pi \ k$.

To simplify the notation we shall call $cos \beta_{j}= \beta_{j}$. Considering only the energy level $E_{0}$,
  \begin{equation}
 d\sigma_{\beta_{0}}=\sigma_{0}/2 \ d\beta_{0}
 \label{eq:one}.
 \end{equation}

 Let us define the ratio of the elastic and the inelastic cross sections as $\lambda (E)=\sigma(E)/s(E)$ , where $\sigma(E)$ is an estimation of the elastic cross section per path length unit and $s(E)$ is the probability per length unit that the electron be inelastically scattered. The isotropy of the elastic cross section makes the expressions angle-independent.
 The ratio $\lambda$ has the advantage that it could be considered as energy independent, making the invariant embedding equations more simple to manage.

In this work, the states ladder model described in ref \cite{Heluani:JAP06} will be used considering five energy states. The energy range of {\it active} electrons (i.e. capable of produce ionization) is divided into intervals or steps in a ladder. The transition (in the ladder) from one step to the following  is described as follows: after an inelastic collision the electrons change their energies by a discrete amount $\delta E$, always degrading its original value, which implies to consider the problem of multiple inelastic collisions as reduced to a problem of only a single effective collision.

\subsection{Invariant embedding equations}

 As usual in IEAM if it is possible to add a new layer of infinitesimal thickness $d\tau$ on the surface of the sample, then we may consider the interval $(0,\tau)$ as a sub-sample lying in the new $(0,\tau+d\tau)$ sample \cite{Bellman:JMP60,Dashen:PRB64,Heluani:NIMB00,Glazov:NIMB07}. Let us introduce the matrix ${\bf R}=(R_{\alpha_{j},\beta_{k}})$ where  $R_{\alpha_{j},\beta_{k}}$ is the probability that an electron that impinges on the sample with polar angle $\alpha_{j}$ and with energy $E_{j}$ leaves the sample with an angle $\beta$ and energy $E_{k}$.
Considering only one energy state, figure 1.b shows all possible trajectories which make contribution to the electron backscattered fraction from a sample of thickness $\tau+d\tau$. Consider as an illustration the probabilities of the occurrence of the paths associated
to the first and second diagrams in figure 1. The probability of occurrence of the first path is the probability that an impinging electron having an energy $E_{0}$ and cosine of the polar incident angle $\alpha_{0}$ passes across the interval $(\tau+d\tau, \tau)$ without suffering elastic collision neither inelastic collision, it is  $(1-s_{0}d\tau/\alpha_{0})(1-\sigma_{0}d\tau/\alpha_{0})$ multiplied by the probability that the electron will be backscattered from a``sub-sample'' of size $\tau$ without loss its energy, it is $R_{\alpha_{0},\beta_{0}}$  and by the probability that leave the sample at $\tau+d\tau$ with an emerging cosine $\beta_{0}$ without suffering collision in the interval $(\tau+d\tau, \tau)$. The probability of the path is written as $(1-s_{0}d\tau/\alpha_{0})^2(1-\sigma_{0}d\tau/\alpha_{0})^2R_{\alpha_{0},\beta_{0}}$. Using the same procedure, the probabilities of the other four paths are written as:
\begin{eqnarray*}
Path \ 2&(1-s_{0}\frac{d\tau}{\alpha_{0}})d\sigma_{\alpha_{0},\delta_{0}}\frac{d\tau}{\alpha_{0}}R_{\delta_{0},\beta_{0},}(1-s_{0}\frac{d\tau}{\beta_{0}})(1-\sigma_{0}\frac{d\tau}{\beta_{0}}), \\
3&(1-s_{0}\frac{d\tau}{\alpha_{0}})(1-\sigma_{0}\frac{d\tau}{\alpha_{0}})R_{\alpha_{0},\epsilon_{0},}(1-s_{0}\frac{d\tau}{\epsilon_{0}})d\sigma_{\epsilon_{0},\beta_{0}}\frac{d\tau}{\epsilon_{0}} \\
4&(1-s_{0}\frac{d\tau}{\alpha_{0}})(1-\sigma_{0}\frac{d\tau}{\alpha_{0}})R_{\alpha_{0},\epsilon_{0},}(1-s_{0}\frac{d\tau}{\epsilon_{0}}))d\sigma_{\epsilon_{0},\varphi_{0}}\frac{d\tau}{\epsilon_{0}}R_{\varphi_{0},\beta_{0},}  \\
5&(1-s_{0}\frac{d\tau}{\alpha_{0}})d\sigma_{\alpha_{0},\beta_{0}}\frac{d\tau}{\alpha_{0}}.
\end{eqnarray*}

 It is important to emphasize the fact that the details of the particle behavior in $(0,\tau)$ are of no interest.  As usual, to obtain $R_{\alpha_{0},\beta_{0}}(\tau + d\tau)$ all probabilities of the paths illustrated in Fig. 1 must be summed taken into account that the probabilities of the trajectories 2, 3 and 4 must be integrated over the cosines $\delta_{0}, \epsilon_{0}$ and $\varphi_{0}$. Then letting $d\tau \rightarrow 0$ to yield to a differential equation for $R_{\alpha_{0},\beta_{0}}(\tau)$ \cite{Bellman:JMP60}. Any other possible paths have a contributions of the order of $(d\tau)^{2}$ and their contribution will vanish in this procedure. The differential equation for $R_{\alpha_{0},\beta_{0}}$, in its simplest form for one level model is written as,

\begin{eqnarray}
\frac{dR_{\alpha_{0},\beta_{0}}(\tau)}{d\tau}=\frac{d\sigma_{\alpha_{0},\beta_{0}}}{\alpha_{0}}+
\int_{\delta_{0}}\frac{d\sigma_{\alpha_{0},\delta_{0}}R_{\delta_{0},\beta_{0}}(\tau)}{\alpha_{0}}
\nonumber  \\
\int_{\varphi_{0}}\int_{\epsilon_{0}}\frac{d\sigma_{\epsilon_{0},\varphi_{0}}R_{\alpha_{0},\epsilon_{0}}(\tau)R_{\varphi_{0},\beta_{0}}(\tau)}{\epsilon_{0}}
\nonumber \\
+\int_{\epsilon_{0}}\frac{d\sigma_{\epsilon_{0},\beta_{0}}R_{\alpha_{0},\epsilon_{0}}(\tau)}{\epsilon_{0}} -R_{\alpha_{0},\beta_{0}}(\tau)[\frac{s_{0}+\sigma_{0}}{\beta_{0}}\frac{s_{0}+\sigma_{0}}{\alpha_{0}}].
\end{eqnarray}

The integro differential Eq. 2 do not have an exact solutions. However, using a a convergence method it is possible to obtain a practical solution to be used in spectroscopy measurements.

Suppose $\tau \approx \infty$ for the trajectories in Fig.1 that involve at least one collision in the differential layer $d\tau$. These are the trajectories 2, 3, 4 and 5 in Fig.1.  Let us introduce the function $r(\alpha_{0},\beta_{0},\infty)$ which  denote the probability of occurrence of these trajectories so that,
\begin{eqnarray}
r(\alpha_{0},\beta_{0},\infty)=\frac{d\sigma_{\alpha_{0},\beta_{0}}}{\alpha_{0}}+
\int_{\delta_{0}}\frac{d\sigma_{\alpha_{0},\delta_{0}}R_{\delta_{0},\beta_{0}}(\infty)}{\alpha_{0}}
\nonumber  \\
\int_{\varphi_{0}}\int_{\epsilon_{0}}\frac{d\sigma_{\epsilon_{0},\varphi_{0}}R_{\alpha_{0},\epsilon_{0}}(\infty)R_{\varphi_{0},\beta_{0}}(\infty)}{\epsilon_{0}}
\nonumber \\
+\int_{\epsilon_{0}}\frac{d\sigma_{\epsilon_{0},\beta_{0}}R_{\alpha_{0},\epsilon_{0}}(\infty)}{\epsilon_{0}}
  \end{eqnarray}
If we now replace the first four terms in Eq.2 by the later expression, an approximate solution for $  R_{\alpha_{0},\beta_{0}}(\tau)$ can be written as,
 \begin{equation}
  R_{\alpha_{0},\beta_{0}}(\tau) = \frac{r(\alpha_{0},\beta_{0},\infty)\beta_{0}\alpha_{0}}{(\alpha_{0}+\beta_{0})(s_{0}+\sigma_{0})}[1-e^{-\tau (\frac{s_{0}+\sigma_{0}}{\beta_{0}}+\frac{s_{0}+ \sigma_{0}}{\alpha_{0}})}],
 \end{equation}

We shall now proceed to derive $r(\alpha_{0},\beta_{0},\infty)$ which makes the approximate solution $R_{\alpha_{0},\beta_{0}}(\tau)$,  Eq.4, to  converge to the  exact solution of Eq.2 for large values of $\tau$.

\subsection{Convergence of the solution}

Let us now proceed to  replace $R_{\alpha_{0},\beta_{0}}(\infty)$ from  Eqs 4. in Eq. 3 and then we proceed to  solve  the following integrals,
 \begin{eqnarray}
 r(\alpha_{0},\beta_{0},\infty)d\beta_{0}=\frac{\sigma_{0}d\beta_{0}}{2\alpha_{0}}\times \{1+\int_{0}^{1} \frac{r(\alpha_{0},\beta_{0},\infty)\beta_{0}\delta_{0}d\delta_{0} }{(\delta_{0}+\beta_{0})(s_{0}+\sigma_{0})}+ \nonumber  \\
  + \alpha_{0}\int_{0}^{1} \frac{d\epsilon_{0} r(\alpha_{0},\beta_{0},\infty)\alpha_{0}}{(\alpha_{0}+\epsilon_{0})(s_{0}+\sigma_{0})}  \nonumber \\
 +\alpha_{0}\int_{0}^{1}\int_{0}^{1} \frac{d\epsilon_{0}d\varphi_{0}r(\alpha_{0},\beta_{0},\infty)^{2}\alpha_{0}\varphi_{0} \beta_{0}}{(\alpha_{0}+\epsilon_{0})(\varphi_{0}+\beta_{0})(s_{0}+\sigma_{0})^{2}}\},
 \end{eqnarray}
where the cosines $\alpha_{0},\ \beta_{0},\  \epsilon_{0}$ and $\varphi_{0}$ are represented in Fig. 1-a. In Eq. 5 we make use of the fact that the integration variables  are the director cosines of the  emerging or of the incident angles, which by symmetry can be exchanged.\\
 Exact evaluation of the integrals in  Eq. 5 is a difficult task, because it is necessary to know explicitly the angular  dependence of $r(\alpha_{0},\beta_{0},\infty)\times \alpha_{0}$.  To proceed, taken into account that the unknown function  is a continuous function of $\alpha_{0},\beta_{0}$  for all angle, we consider that the value of $r(\alpha_{0},\beta_{0},\infty) \times \alpha_{o}$ in the integrands of Eq. 5 is nearly independent of the director cosines and equal to a constant which must be estimated using physical arguments. One method could be to use the mean-value theorem. However, in this work we consider an {\it effective} value $r^*$  that could be calculated from the following procedure:  substitute $r(\alpha_{0},\beta_{0},\infty)\times \alpha_{o}$ by the unknown constant $r^{*}$ in the arguments of the integrals in Eq.5 and then evaluate the right hand  member. The result for $r(\alpha_{0},\beta_{0},\infty)$ is,
 \begin{eqnarray}
 \alpha_{0}r(\alpha_{0},\beta_{0},\infty)=\sigma_{0}/2+\frac{r^* \beta_{0} \sigma_{0}}{2(s_{0}+\sigma_{0})}Ln(1+
\frac{1}{\beta_{0}}) \nonumber \\
+\frac{\sigma_{0}r^* \alpha_{0}}{2(s_{0}+\sigma_{0})}Ln(1+\frac{1}{\alpha_{0}})
+\frac{(r^*)^{2}\alpha_{0} \sigma_{0}\beta_{0}}{2(s_{0}+\sigma_{0})}\times \nonumber
\\ Ln(1+\frac{1}{\alpha_{0}})Ln(1+\frac{1}{\beta_{0}}).
 \end{eqnarray}
Now we make use of this later results to find the value of $r^{*}$ that satisfies the following conditions (considering the integrals in Eq. 5) ,
\begin{equation}
\int_{0}^{1}\int_{0}^{1} \frac{r(\alpha_{0},\beta_{0},\infty)\alpha_{0}\beta_{0}d\alpha_{0}d\beta_{0} }{(\alpha_{0}+\beta_{0})(s_{0}+\sigma_{0})}= \int_{0}^{1}\int_{0}^{1} \frac{r^{*}\beta_{0}d\alpha_{0}d\beta_{0}}{(\alpha_{0}+\beta_{0})(s_{0}+\sigma_{0})}
\end{equation}
Then, using (6) and (7) the final expression for $R_{\alpha_{0},\beta_{0}}(\tau=\infty)$  in Eq.4 is,
 \begin{eqnarray}
R_{\alpha_{0},\beta_{0}}(\infty)=\frac{C}{(\lambda+1)(\alpha_{0}+\beta_{0})}\times\{\frac{\lambda\Gamma_{1}\beta_{0}}{2}\nonumber \\
+\frac{2\Gamma_{1}^{2}\alpha_{0}\beta^{2}_{0}}{\lambda}Ln(1+1/\beta_{0})Ln(1+1/\alpha_{0})\nonumber \\+\beta^{2}_{0}Ln(1+1/\beta_{0})+\Gamma_{1}\alpha_{0}\beta_{0}Ln(1+1/\alpha_{0})\}.
\end{eqnarray}
where $\Gamma_{1}= \lambda +2-2\sqrt{\lambda+1}$, and C is a normalization constant which considers the conservation of the particle fluxes.

Another useful function in characterization techniques is the fraction of absorbed electrons. Defining the matrix ${\bf A}=(A_{\alpha_{0}}(\tau))$, where $A_{\alpha_{0}}(\tau)$ is the probability that an electron that impinges in the sample with cosine $\alpha_{0}$ and with energy $E_{0}$ becomes {\it deactivated} (i.e., it loses its capacity to produce ionization) inside the sample. This probability can be obtained using the convergence method outlined above.  To do this, we have to construct diagrams similar to that  shown in Fig. 1-b. (see Ref \cite{Heluani:NIM02} for details). The resulting equation is:
  \begin{eqnarray}
\frac{dA_{\alpha_{0}}(\tau)}{d\tau}=\frac{d\sigma_{\alpha_{0},\beta_{0}} A_{\beta_{0}}(\tau)}{\alpha_{0}}+ \frac{s_{0}}{\alpha_{0}}+ \int_{\delta_{0}}\frac{s_{o} R_{\alpha_{0},\beta_{0}}(\tau)}{\delta_{0}}+\nonumber \\ \int_{\epsilon_{0}}\int_{\varphi_{0}}\frac{d\sigma_{\epsilon_{o}\varphi_{0}} R_{\alpha_{0},\epsilon_{0}}(\tau)A_{\varphi}(\tau)}{\epsilon_{0}}-
A_{\alpha_{0}}(\tau)[\frac{\sigma_{0}}{\beta_{0}}+\frac{s_{0}}{\alpha_{0}}].
  \end{eqnarray}
The solutions are:
  \begin{eqnarray}
 A_{\alpha_{0}}(\tau)& =& A_{\alpha_{0}}(\infty)[1-e^{-\frac{\tau}{\alpha_{0}} (s_{0}+\sigma_{0})}],\nonumber \\
                  &  & A_{\alpha_{0}}(\infty)=\frac{a(\infty)\alpha_{0}}{s_{o}+\sigma_{0}}.
 \end{eqnarray}
An expression for $a(\infty)$  is easy to obtain using the same procedure as for $r(\infty)$
\begin{equation}
 A_{\alpha_{0}}=\frac{C}{(\lambda+1)}\times[1+\Gamma_{2}+ \frac{2\Gamma_{1}\alpha_{0}}{\lambda}Ln(1+1/\alpha_{0})(1+\Gamma_{2})].
  \end{equation}

  \subsection{Numerical evaluation}
On the basis of the  procedure described above a set of closed expression for $R_{\alpha_{0};\beta_{k}}(\tau ),\  k= 1, 2, 3 ,4$ and $5$, is obtained using the states ladder model described in Ref. 14.  The algebra is cumbersome but using  a standard PC the calculation  can be performed in a few minutes using any mathematical software. The method was applied to a number of elements of representative atomic numbers to illustrate its qualitative behavior.

 The numerical results depend on the approximations used to estimate the cross sections for the scattering processes involved. Although any approximation could be used in IIAM equations, in this work the Rutherford elastic cross section is used \cite{William:RMP63}. The inelastic cross section is estimated from Bethe theory \cite{Bethe:PAP38}. Then, the ratio $\lambda$ could be expressed as:
\begin{equation}
\lambda = 3.96 \ 10^{-20} \frac{[Ln(\frac{E}{\nu})^{2}-Ln(\frac{E}{I})^{2}]}{E}\frac{\rho N_{A}}{A}.
\end{equation}
where $\nu$ is the minimum value of energy loss in each inelastic collision, $I$ is the ionization energy of  the  atoms. $\rho$, $N_{A}$ and  $A$ are the density, the Avogadro constant and the atomic number respectively.

The  backscattered fraction $\eta$, calculated as a function of the atomic number Z and the  energy distribution for various materials are show in Figure 2a and 2b, respectively. Experimental data in figure 2a were taken from {\it Database of electron-Solid Interaction} \cite{Joy:USA}.
 Figure 2a shows theoretical calculation considering $\nu$ as a fitting  parameter in the ratio $\lambda$ (or in the cross section expression). The best fit is obtained with $\nu$ close to 4.25 $\times$10$^{-11}$KeV.
 The theoretical  results lead to reasonably accurate agreement with the experimental data  for the atomic number dependence. However, in the case of the energy spectrum for the backscattered electrons, our theoretical model predicts a maximum at the incident energy E0. Actually, in experimental results the maximum in the spectrum occurs at energies slightly below E0 (see Ref. \cite{Matsukawa:JPD74}). The discrepancy could be explained taking into account that our model considers the possibility of perfect elastic collisions, while actual collisions occur with certain amount of energy loses.

  Even though the limitations of the convergence approach to describe the maximum expected in Fig.3a, Eq. 2 was derived from first principles, straight from the model of the physical process. Therefore, it does not contain the strong physical hypothesis assumed in most of the models employed in different software packages for EPMA, namely, that the backscattered electron trajectories and their energy loss mechanisms inside the solid, are independent of the trajectories of the {\it deactivated} (absorbed) electrons.

 A qualitative behavior of the theoretical results for the fraction of absorbed electrons as a function of incident angle is presented in figure 3. In agreement with experimental results, the maximum occurs at normal incidence of the electron beam.

\section{Summary}

The aim of this work is to report a new approach to solve the functional equations of IE that describes physical processes in spectroscopic techniques. With this approach it is possible to obtain, with relative facility, analytical expressions of accessible treatment which could be useful to the experimental investigators to interpret their
results. These expressions also facilitate the evaluation of such parameters like cross sections, attenuation factors, etc. In the present work it we have dealt with the calculation of the absorbed and backscattered electronic fractions, improving previous models and thus allowing a closer approximation to the real phenomenon. This approach offers an different point of view for the study of the already mentioned
physical parameters. At present there are calculations in progress of other parameters of interest in spectroscopy, such as the k reasons \cite{Heluani:XRS05} applying the convergence method. This study will allow an independent validation of the model and method previously described.

\begin{acknowledgments}
his work  was partially supported by CIUNT under Grants
26/E439 by ANPCyT-PICTR 20770 and 35682.
\end{acknowledgments}

\bibliographystyle{prsty}

\vfill
\newpage
\begin {center}
{\bf Figure captions}
\end {center}
\noindent Fig. 1: a) Angles defined for equation 3.  b) Five trajectories which make contributions to the backscattered electron fraction in a one-state model. Here $\alpha$, $\beta$ etc indicate the cosines of the corresponding angles.
\vskip 1cm
\noindent Fig. 2: a) Backscattered electrons fraction, $\eta$, as a function of atomic number. Theoretical values, color points. Experimental values, crosses. b)Energy spectrum of backscattered electrons from samples of different atomic number Z.
\vskip 1cm
\noindent Fig. 3:The fraction of absorbed electrons as a function of incident angle.



\newpage
\begin{figure}
\includegraphics{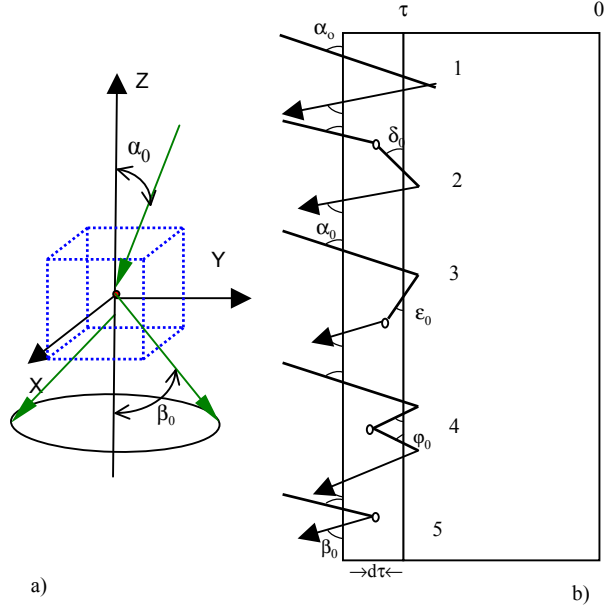}
 \caption{\small{\it{:a) Angles defined for equation 3.  b) Five trajectories which make contributions to the backscattered electron fraction in a one-state model. Here $\alpha$, $\beta$ etc indicate the cosines of the corresponding angles.}}}
\end{figure}
\newpage
\begin{figure}
\includegraphics{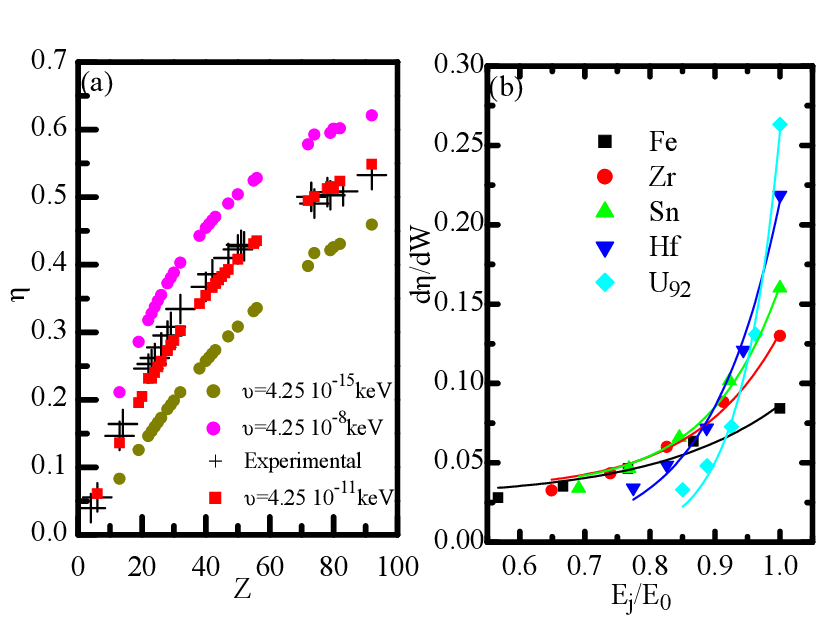}
 \caption{\small{\it{ a) Backscattered electrons fraction, $\eta$, as a function of atomic number. Theoretical values, color points. Experimental values, crosses. b)Energy spectrum of backscattered electrons from samples of different atomic number Z.}}}
     \end{figure}
\newpage
\begin{figure}
\includegraphics{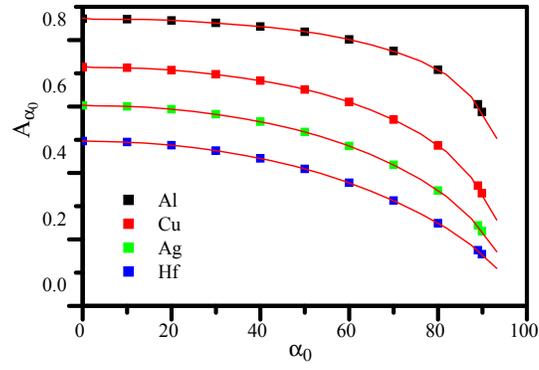}
 \caption{\small{\it{The fraction of absorbed electrons as a function of incident angle.}}}
     \end{figure}
\end{document}